\documentclass[12pt]{iopart}
\usepackage{graphicx}
\usepackage{siunitx}
\usepackage{citesort}

\newcommand{\appropto}{\mathrel{\vcenter{
  \offinterlineskip\halign{\hfil$##$\cr
    \propto\cr\noalign{\kern2pt}\sim\cr\noalign{\kern-2pt}}}}}

\begin{document}
\title[On the control of the spatio-temporal properties of biphoton states]{Where are photons created in parametric down conversion? On the control of the spatio-temporal properties of biphoton states}
\author{  A B\"{u}se$ˆ{1,2,3}$,   N Tischler$ˆ{1,2,3}$, M L Juan$ˆ{1,2}$ and G Molina-Terriza$ˆ{1,2}$\footnote{Corresponding author}}
\address{$ˆ1$ Centre for Engineered Quantum Systems, Macquarie University, NSW 2109, Australia}
\address{$ˆ2$ Department of Physics \& Astronomy, Macquarie University, NSW 2109, Australia}
\address{$ˆ3$ Joint first authors}

\ead{gabriel.molina-terriza@mq.edu.au}
\begin{abstract}
In spontaneous parametric down-conversion photons are known to be created coherently and with equal probability over the entire length of the crystal. Then, there is no particular position in the crystal where a photon pair is created. We make the seemingly contradictory observation that we can control the time delay with the crystal position along the propagation direction. We resolve this contradiction by showing that the spatio-temporal correlations critically affect the temporal properties of the pair of photons, when using a finite detector size. We expect this to have important implications for experiments that require indistinguishable photons.

\end{abstract}
\maketitle

\section{Introduction}

The nonlinear process of spontaneous parametric down-conversion (SPDC), in which a small fraction of photons from an intense pump beam is converted into pairs of lower energy photons, is currently the  predominant method for producing photon pairs \cite{Clausen2011,Kim2001} and heralded single photons \cite{Fasel2004}. These are required for several applications of quantum optics, including quantum computation \cite{Walther2005}, quantum metrology \cite{Sergienko2003,Afek2010}, quantum communication \cite{Trifonov2005,Ursin2007}, and fundamental tests of quantum mechanics \cite{Dada2011,Tittel2008}. Such diversity of applications has been made available through the possibility to tailor properties of the down-converted photons, which depend on the polarisation, spectral/temporal, and spatial degrees of freedom. Photon pairs can be entangled in each of, and even between, these degrees of freedom \cite{Wong2006,Kwiat1993,Molina-Terriza2007,Barreiro2005,Nagali2009}. A lot of progress has been made in developing ways to achieve frequency or momentum correlated, uncorrelated, and anticorrelated states \cite{Grice2011,Hendrych2007,Rangarajan2011,Shimizu2009,Yun2012,Valencia2007}. Temporal indistinguishability is a common requirement for quantum optics experiments, especially in applications that rely on quantum interference. In the case of type II SPDC sources this is often achieved by birefringent compensation \cite{Search2003,Tittel2008,Wong2006}.

The photons from SPDC are generally well understood and known to be created coherently and, in the absence of pump losses, with equal probability over the length of the crystal \cite{Search2003}. In theory, the number of down-converted photons depends on the total pump power and, unlike second harmonic generation, is independent of the pump intensity distribution for a fixed pump divergence and power. This is basically due to the low efficiency of the nonlinear effect which leaves the pump beam essentially undepleted. The total number of photon pairs created is then independent of the location of the pump beam relative to the crystal \cite{DiLorenzoPires2011,Bennink2006}. On the other hand, under realistic experimental conditions, the number of detected photons will strongly depend on the collection optics and detector size.

Here we report the measurement of a time delay between the photon pair in type II SPDC that depends on the position of the crystal. We will experimentally and theoretically show that this effect is not due to a crystal localized photon creation or detection. Instead, the cause of the tunable time delay can be attributed to the particular spatio-temporal structure of the biphoton wavefunction. The frequency momentum correlations induced by the SPDC process are akin to those present in localized waves \cite{Saari2004}. We show that the collection lens introduces a spatially dependent quadratic phase, which through the spatio-temporal correlations, becomes a temporal delay between the photon pairs arriving at the detectors. 

When concentrating on either spectral or spatial correlations between signal and idler photons, spatio-temporal correlations are often considered a nuisance and are avoided through strong filtering in the other degree of freedom. However, without such filtering, pronounced correlations exist between frequency and transverse momentum \cite{Osorio2008}. These correlations are currently subject of intense research \cite{Gatti2009,Brambilla2010,Gatti2012,Gatti2014}.


As explained in more detail in section \ref{Experimental Method}, a popular SPDC source consists of a focused pump beam, a nonlinear crystal designed for type II collinear down-conversion, and a lens to collect the photon pairs. The usual protocol for setting the relative positions of the crystal and collection lens is to centre the pump focal point in the middle of the crystal and then set the collection lens to maximise the count rates. Hence, the positions of the crystal and collection lens are fixed. One of the few investigations into the effect the crystal position has on the two-photon state was done by Di Lorenzo Pires \emph{et al.}, who studied the near-field intensity patterns in type I emission \cite{DiLorenzoPires2011}. In the following, we also allow the crystal and collection lens positions to be variable and find that their distance has a strong impact on the delay between signal and idler. We establish this for single-mode fibre and free-space detection through both measurements and theory.

\section{Theoretical Model}

We will start by providing a complete theoretical model of the frequency and momentum correlations of the biphoton wavefunction and their effect in relevant experimental conditions.  We therefore set out to calculate the number of coincidence counts as a function of time delay between signal and idler photons from type II down-conversion, arriving at the detector for two different detection schemes: single-mode fiber collection and free space detection. Since we regard the pump focal position and distance between crystal and collection lens as free parameters, they are to appear explicitly in the calculation.

Let us begin with the biphoton wavefunction at the crystal exit facet, which already contains the dependence on the pump focal position. Throughout this work, we will model our experiment with a monochromatic pump beam, which means that signal and idler frequencies add up to a fixed pump frequency. The biphoton wavefunction is
\begin{equation} \label{Psi_general}
|\Psi \rangle = \int \mathrm{d}\mathbf{q}_s \mathrm{d}\mathbf{q}_i \mathrm{d} \omega_s ~\Phi(\mathbf{q}_s,\mathbf{q}_i,\omega_s,\omega_i) \hat{a}^{\dagger}(\mathbf{q}_s,\omega_s,\sigma_s)  \hat{a}^{\dagger}(\mathbf{q}_i,\omega_i,\sigma_i) |0\rangle ,
\end{equation} 
where $s$ and $i$ label signal and idler photons, $\mathbf{q}$ is the transverse wavevector, $\omega$ the angular frequency, and $\sigma$ the polarisation. $\hat{a}^{\dagger}(\mathbf{q},\omega,\sigma)$ is the photon creation operator for a photon mode characterized by transverse wavenumber, frequency, and polarisation. Throughout this work, our integrals are implied to be over all possible values of the integration variables, unless specified otherwise.  $\Phi(\mathbf{q}_s,\mathbf{q}_i,\omega_s,\omega_i)$ is the biphoton mode function, which  takes on the form
\begin{eqnarray} \label{Phi_crystal_end}
\Phi(\mathbf{q}_s,\mathbf{q}_i,\omega_s,\omega_i)&\propto& \mathrm{sinc}\left(\frac{\Delta k_z(\mathbf{q}_s,\mathbf{q}_i,\omega_s,\omega_i)L}{2}\right)\mathrm{exp}\left(-\frac{w^2  |\mathbf{q}_{s}+\mathbf{q}_{i}|^2}{4}\right) \nonumber \\
 &&\times \mathrm{exp}(i~k_{pz}(z_c-z_{foc}(z_c)))\nonumber \\
&&\times \mathrm{exp}(i~(k_{sz}(\omega_s,\mathbf{q}_s)+k_{iz}(\omega_i,\mathbf{q}_i))~L/2). 
\end{eqnarray} 
Here $L$ is the crystal length, $w$ the pump beam waist,  $\Delta k_z$ the longitudinal wave vector mismatch $k_{pz}-k_{sz}-k_{iz}-\frac{2\pi}{\Lambda}$ and $\Lambda$ the poling period of the crystal. $\omega_i$ is not an independent variable as it is given by $\omega_i=\omega_p-\omega_s$, but we do keep it in the expressions for the sake of clarity. Our main experiment consists in longitudinally displacing the crystal, and to model this process, we need to consider the effect this has on the pump focal position. The geometry is illustrated in \mbox{figure \ref{crystal_geometry}}. $z_c$ denotes the position of the crystal centre and $z_{foc}(z_c)$ the focus of the pump beam, both  relative to the position at which the focus coincides with the crystal centre (i.e. $z_{foc}=0$ when $z_c=0$). Clearly, in the commonly assumed case of the pump focal point lying at the centre of the crystal, the $\mathrm{exp}(i~k_{pz}(z_c-z_{foc}(z_c)))$ term disappears. Assuming the pump beam is paraxial, its focal position in the laboratory frame is given by:

\begin{figure}[t]
	\centering
		\includegraphics[width=1\textwidth]{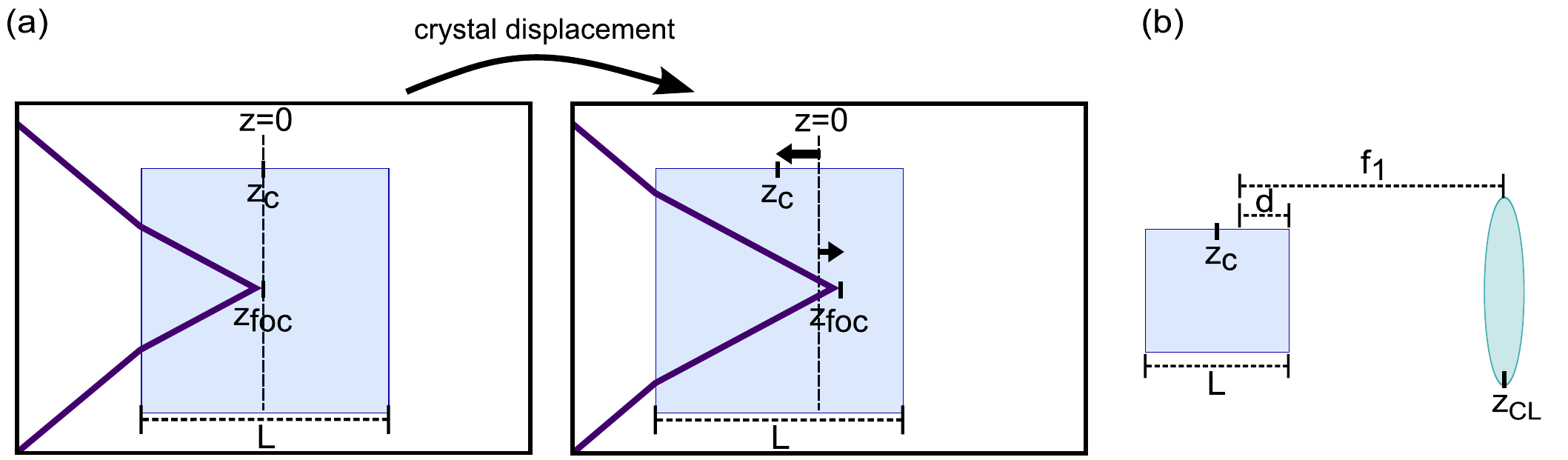}
	\caption{(a) Longitudinal shift of the pump focus caused by a translation of the crystal. $z=0$ is the reference position at which the pump focus coincides with the crystal centre. $z_c$ is the position of the crystal centre and $z_{foc}$ the position of the pump focus. $L$ is the length of the crystal. This configuration is similar to the one used in \cite{DiLorenzoPires2011}. (b) Illustration of our geometry after the crystal. $z_{CL}$ is the position of the collection lens which has a focal length $f_1$. $d$ is the distance between a point one focal distance before the collection lens, and the crystal end.} 
	\label{crystal_geometry}
\end{figure}

\begin{equation}
z_{foc}(z_c)=\left\{
     \begin{array}{lr}
       \frac{L}{2}-\frac{L}{2n_p} & : z_c\le-\frac{L}{2n_p} \\
       z_c-n_p z_c & : -\frac{L}{2n_p}<z_c<\frac{L}{2n_p} \\
       -\frac{L}{2}+\frac{L}{2n_p} & : z_c\ge\frac{L}{2n_p},
     \end{array}
   \right.
\end{equation}
where $n_p$ is the refractive index of the crystal experienced by the pump beam. The crystal positions $|z_c|=\frac{L}{2n_p}$ correspond to the pump focus lying at one of the crystal facets. Within the crystal, the pump focal position shifts in the opposite direction to the crystal movement. Once shifted out of the crystal in either direction, it no longer depends on the crystal position. As mentioned earlier, from here there are two possible treatments of the spatial degree of freedom, depending on the detection scheme used. 

For single-mode fibre detection, a projection into a Gaussian detection mode is performed. We can do this at the crystal end facet where we have the expression for the biphoton wavefunction (\ref{Psi_general},\ref{Phi_crystal_end}). The Gaussian detection mode is 
\begin{eqnarray}
G_{spa}(\mathbf{q})_{z=z_c+L/2}&=&\frac{w_f}{\sqrt{2\pi}}~\mathrm{exp}\left(-\frac{w_f^2 |\mathbf{q}|^2}{4}\right)\mathrm{exp}\left(-i\frac{|\mathbf{q}|^2 d}{2 k_{air}(\omega) } \right),
\end{eqnarray}
where $w_f$ is the detection mode beam waist, and $k_{air}(\omega)=\frac{\omega}{ c}$ is the wavenumber of the detected photon in air, $c$ being the speed of light in vacuum. 
The Gaussian mode is effectively characterised by two things, the distance of the crystal end from where the detection focal point would be in the absence of the crystal, $d=\left(f_1 - (z_{CL}-z_c-\frac{L}{2})\right)$, and the beam waist, $w_f$. The optimal detection beam waist depends on the pump beam waist among other things, and a body of work on this subject is available in the literature \cite{Bennink2010,Smirr2013,Guerreiro2013}. Usually, the detection focal point is assumed to lie at the centre of the crystal, coinciding with the pump focal point, and this imposes a particular distance from the crystal end to where the detection focal point would be without the crystal, $\left(\frac{L/2}{n}\right)$. The position of the collection lens for this case is $z_{CL}=\frac{L}{2}-\frac{L/2}{n}+f_1$. As we allow for longitudinal translations of the collecting lens, $z_{CL}$ will remain a parameter in our calculations. After projection of the photons into Gaussian detection modes with equal detection beam waists for signal and idler, the wavefunction reads:
  \begin{eqnarray}
&|\Psi_{SMF} \rangle =& \int \mathrm{d} \omega_s ~\Phi_{SMF}(\omega_s,\omega_i) \hat{a}^{\dagger}(\omega_s,\sigma_s)  \hat{a}^{\dagger}(\omega_i,\sigma_i) |0\rangle ,\nonumber \\ 
 \Phi_{SMF}(\omega_s,&\omega_i) =& \int \mathrm{d}\mathbf{q}_s \mathrm{d}\mathbf{q}_i \Phi(\mathbf{q}_s,\mathbf{q}_i,\omega_s,\omega_i) G^*_{spa}(\mathbf{q}_s) G^*_{spa}(\mathbf{q}_i).
\end{eqnarray}
In order to calculate the time delay distribution between the arrival times of signal and idler photons, the following expression is evaluated:
\begin{eqnarray} \label{SMFCoinc}
R_{coinc,SMF}(\tau)&\propto&|\langle 0|\hat{E}_i^{(+)}(t-\tau/2) \hat{E}_s^{(+)}(t+\tau/2) |\Psi_{SMF}\rangle |^2,
\end{eqnarray}
where the operator $\hat{E}^{(+)}_j(t)$ is proportional to $\int \mathrm{d}\omega ~\mathrm{exp}(-i\omega t)\hat{a}(\omega,\sigma_j)$. 
The reason $t$ does not appear on the left-hand-side is that for a monochromatic pump beam, the quantity is independent of the mean time; it only depends on the time difference.

In contrast to the single-mode fibre case where only one spatial mode is relevant at the detector, for free-space detection, the counts as a function of time delay need to be calculated for all pairs of points on the detector surfaces, and subsequently integrated over all such available pairs:
\begin{eqnarray} \label{FSCoinc}
R_{coinc,FS}(\tau)&=&\int_{\mathrm{A_{det}}} \mathrm{d} \mathbf{r}_s \mathrm{d} \mathbf{r}_i R_{coinc,PP}(\mathbf{r}_s,\mathbf{r}_i,\tau)\nonumber \\
&\propto&\int_{\mathrm{A_{det}}} \mathrm{d} \mathbf{r}_s \mathrm{d} \mathbf{r}_i |\langle 0|\hat{E}_i^{(+)}(\mathbf{r}_i,t-\tau/2) \hat{E}_s^{(+)}(\mathbf{r}_s,t+\tau/2) |\Psi\rangle |^2
\end{eqnarray}
The electric field operator at the detector plane can be related to the annihilation operator at the crystal exit facet using a thin lens model from Fourier optics and the Fresnel approximation:
\begin{eqnarray} \label{EfieldRelation}
\hat{E}(\mathbf{r},t)_{z=z_{det}}&=&\int \mathrm{d}\omega~ \mathrm{d}\mathbf{q}  ~\mathrm{exp}\left(-i \frac{f_1}{f_2}\mathbf{r}\cdot \mathbf{q}-i\omega t\right)\nonumber \\
&&\times \mathrm{exp}\left(-ik_{air,z} d\right)  \hat{a}(\mathbf{q},\omega)_{z=z_c+L/2} \nonumber \\
&\approx&\int \mathrm{d}\omega~ \mathrm{d}\mathbf{q}  ~\mathrm{exp}\left(-i \frac{f_1}{f_2}\mathbf{r}\cdot \mathbf{q}-i\omega t\right)\nonumber \\
&&\times~\mathrm{exp}\left(i\left(-k_{air}(\omega)+\frac{|\mathbf{q}|^2}{2k_{air}(\omega)}\right) d\right)\hat{a}(\mathbf{q},\omega)_{z=z_c+L/2}, 
\end{eqnarray}
where $f_1$ and $f_2$ are the focal lengths of the collection lens and of the focusing lenses in front of the detectors, respectively. For each pair of points that contribute to the integral in equation (\ref{FSCoinc}) and for the expression from the Gaussian detection scheme (\ref{SMFCoinc}), the coincidence counts as a function of time delay can be cast in the form of
\begin{eqnarray}\label{Integral}
R_{coinc}(\tau)&\propto&\bigg|\int \mathrm{d}\mathbf{q}_s \mathrm{d}\mathbf{q}_i \mathrm{d}\Omega ~\exp\left(i (k_{sz}(-\Omega,\mathbf{q}_s)+k_{iz}(\Omega,\mathbf{q}_i))~L/2 \right)\nonumber \\
 &&\times \mathrm{exp}\left(id\left(\frac{|\mathbf{q}_{s}|^2 }{2 k_{air}(-\Omega) }+\frac{|\mathbf{q}_{i}|^2}{2 k_{air}(\Omega) }\right)\right)\nonumber \\
&&\times \mathrm{sinc}\left( \Delta k_z\frac{L}{2} \right) g(\mathbf{q}_s,\mathbf{q}_i,z_{c})~\mathrm{exp}(i \Omega\tau)~\bigg|^2.
\end{eqnarray}

\begin{figure}[t]
	\centering
		\includegraphics[width=0.8\textwidth]{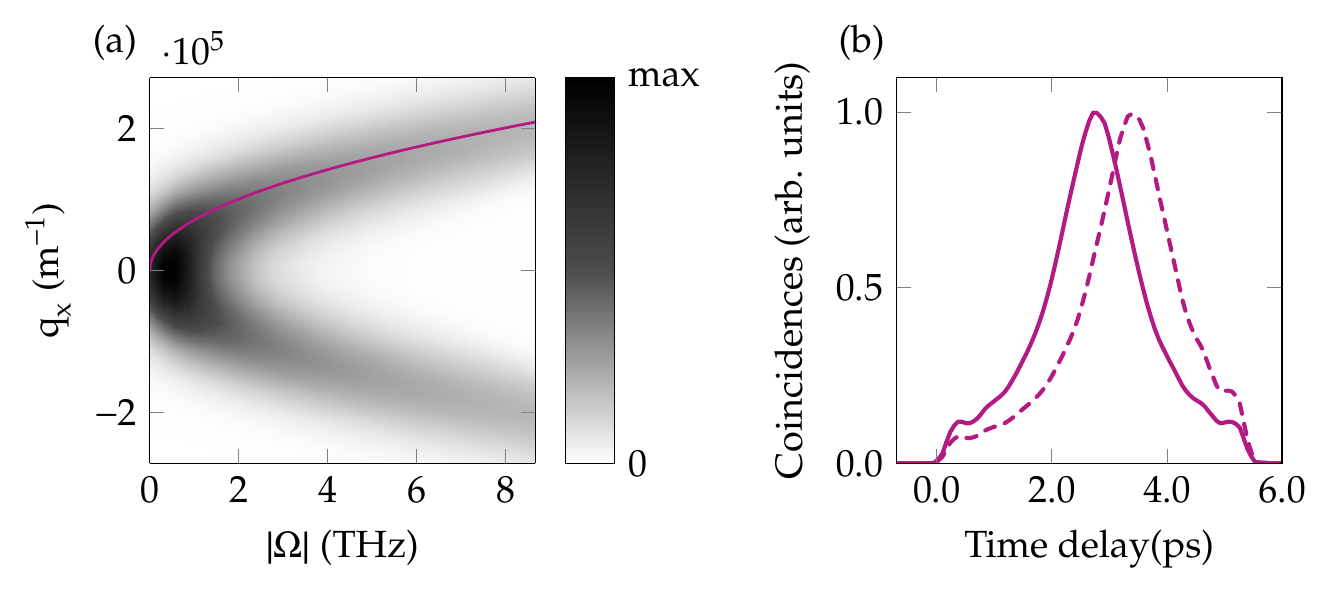}
	\caption{(a) Calculated spatio-temporal correlations of the SPDC wavefunction for experimentally relevant parameters. What we depict is proportional to the probabilities of different $q_x$, $\Omega$ values for a photon with $q_y=0$, after having traced out the other photon of the pair. The overlaid line illustrates the linear dependence between $\Omega$ and $|\mathbf{q}|^2$. (b) Simulated time delay distribution between signal and idler photons for single-mode fibre detection, with the crystal in the central position (solid line) and shifted by \SI{1}{mm} (dashed line).}
	\label{fig:spatio-temporal-correlations}
\end{figure}

\noindent $\Omega$ is defined as $\Omega=\omega_i-\omega_p/2$, and the shorthand of $k_{air}(\Omega)$ means $k_{air}(\omega)=k_{air}(\omega_p/2+\Omega)$. $g(\mathbf{q}_s,\mathbf{q}_i,z_{c})$ incorporates the remaining terms and depends on the type of detection and on the crystal position through the pump focal position. The term that models a change in the distance between crystal and collection lens is $\mathrm{exp}\left(id\left(\frac{|\mathbf{q}_{s}|^2 }{2 k_{air}(-\Omega) }+\frac{|\mathbf{q}_{i}|^2}{2 k_{air}(\Omega) }\right)\right) $. What impact does such a transverse momentum-dependent phase have on the time delay? A potential impact must be mediated by spatio-temporal correlations. Hence, a way to develop an understanding is by considering the form of spatio-temporal correlations imposed by the phase-matching conditions of SPDC. 

To gain an intuitive picture of the system, in the following we develop a toy model by introducing approximations. We will, however, come back to the exact equations (\ref{SMFCoinc}) and (\ref{FSCoinc}) for the simulation which we compare to the experimental results in section \ref{results}. To learn about the spatio-temporal correlations, we perform a multivariate Taylor approximation of $k_{sz}$, $k_{iz}$ and $\Delta k_z$, about the collinear degenerate case up to the first nonzero terms. The reference, for which we take $k_{pz}-k_{sz}-k_{iz}-\frac{2\pi}{\Lambda}=0$, is therefore at $\Omega=0, ~T=T_0, ~\mathbf{q}_s=\mathbf{q}_i=\mathbf{0}$ ($T$ is the crystal temperature). For now, we will further simplify our analysis by considering the case of a plane wave pump beam with $\mathbf{q}_p=\mathbf{0}$ for our toy model. This imposes $-\mathbf{q}_s=\mathbf{q}_i\equiv \mathbf{q}$, and hence $ |\mathbf{q}_s|^2=|\mathbf{q}_i|^2= |\mathbf{q}|^2$ which will allow us to immediately draw some interesting conclusions: 
\begin{eqnarray} \label{Taylor}
\Delta k_z&\approx&\Omega D + (T-T_0) E+ \frac{|\mathbf{q}|^2}{2k_{s}(0)}+ \frac{|\mathbf{q}|^2}{2k_{i}(0)},
\end{eqnarray}
where $D=\left(\frac{\partial k_s}{\partial\Omega}-\frac{\partial k_i}{\partial\Omega}\right)$ and $E=\left(\frac{\partial k_p}{\partial T}-\frac{\partial k_s}{\partial T}-\frac{\partial k_i}{\partial T}\right)$. From now on, all derivatives are evaluated at the reference settings mentioned above. With very long crystals and plane wave pump beam, the photons would be generated only in the perfect phase matching condition, $\Delta k_z=0$. We see that this case entails a linear dependence between $\Omega$ and $|\mathbf{q}|^2$, indicated by the overlaid line in figure \ref{fig:spatio-temporal-correlations} (a) and also shown in reference \cite{Brambilla2010}. As a result, the q-dependent phase term induced by the coupling lens causes a shift of the time delays which is proportional to the change in distance between the crystal and the coupling lens. This is the main mechanism we have identified causing the observed crystal dependent time delay of the emitted photons. We would like to note that the absence of the linear term in  momentum in Eq. (\ref{Taylor}) is due to the fact that in our configuration the photons propagate along one of the optical axes; in other configurations there can be a linear dependence in the momentum \cite{Brida2007}.

Let's go back to the finite crystal model where the sinc function actually results in a spread of $|\mathbf{q}|^2$ values for a given $\Omega$ value. To account for this spread and deepen our first grasp of the underlying physics, let us analyse (\ref{Integral}), again using the Taylor approximation of the longitudinal wavevector mismatch (\ref{Taylor}).
\begin{eqnarray}\label{Integral_analysis1}
R_{coinc}(\tau) & \appropto & \bigg|\int\mathrm{d}\mathbf{q}\mathrm{d}\Omega\,\mathrm{exp}\left(i\frac{{L}}{2}\left(k_{s}(0)+k_{i}(0)+\left(T-T_{0}\right)\left(\frac{{\partial k_{s}}}{\partial T}+\frac{{\partial k_{i}}}{\partial T}\right)-\Omega D\right)\right)\nonumber ~~~~~~~~~~\\
 &  &\times \mathrm{{exp}}\left(id\left(\frac{{|\mathbf{q}|^{2}}}{2k_{air}(-\Omega)}+\frac{{|\mathbf{q}|^{2}}}{2k_{air}(\Omega)}\right)\right)\nonumber \\
 &  &\times \mathrm{{sinc}}\left(\left(\Omega D+\left(T-T_{0}\right)E+\frac{{|\mathbf{q}|^{2}}}{2k_{s}(0)}+\frac{{|\mathbf{q}|^{2}}}{2k_{i}(0)}\right)\frac{L}{2}\right)g(\mathbf{q})~\mathrm{exp}(i\Omega\tau)\bigg|^{2}
\end{eqnarray}
Since we are assuming a plane wave pump and therefore $\mathbf{q}_s=-\mathbf{q}_i$, the integrals over signal and idler transverse momenta were replaced by an integral over one transverse momentum. Importantly, note that the dependence of $g$ on the crystal position, $z_c$, has also dropped out for a plane wave pump. We can now proceed with equation (\ref{Integral_analysis1}) as follows: Since we are taking the modulus, we can remove the phase that is independent of the integration variables. We next evaluate the integral over $\Omega$, which is an inverse Fourier transform of a sinc function. Then, the functions that contain no dependence on $\mathbf{{q}}$ can be taken outside of the remaining integral. This leaves us with
\begin{eqnarray}
R_{coinc}(\tau) & \appropto & \bigg|\frac{{2\pi}}{DL}\mathrm{{rect}}\left(\frac{{1}}{DL}\left(\tau-D\frac{{L}}{2}\right)\right)\nonumber \\
 &  &\times \int\mathrm{d}\mathbf{q}\,\mathrm{exp}\left(id\left(\frac{{|\mathbf{q}|^{2}}}{2k_{air}(-\Omega)}+\frac{{|\mathbf{q}|^{2}}}{2k_{air}(\Omega)}\right)\right)\nonumber \\
 &  & \times\mathrm{{exp}}\left(-i\left(\tau-D\frac{{L}}{2}\right)\frac{{1}}{D}\left(\frac{{|\mathbf{q}|^{2}}}{2k_{s}(0)}+\frac{{|\mathbf{q}|^{2}}}{2k_{i}(0)}\right)\right)g(\mathbf{q})\bigg|^{2}
\end{eqnarray}
At this point, making the approximation of $k_{s}(0)\approx k_{i}(0)\approx nk_{air}(\Omega)\approx nk_{air}(-\Omega)\approx nk_{air}(0)$ shows us the primary effect of changing the distance between crystal
and collection lens:
\begin{eqnarray}\label{Integral_analysis}
R_{coinc}(\tau)  & \appropto & \frac{{2\pi}}{DL}\mathrm{{rect}}\left(\frac{{1}}{DL}\left(\tau-D\frac{{L}}{2}\right)\right) \nonumber \\ 
&&\times \bigg|\int\mathrm{d}\mathbf{q}\,\mathrm{{exp}}\left(\frac{{-i|\mathbf{q}|^{2}}}{nDk_{air}(0)}(\tau-\tau_0)  \right) g(\mathbf{q})\bigg|^{2}
\end{eqnarray}
where $\tau_0=DL/2+nD(f_{1}-z_{\mathrm{CL}}+z_{c}+L/2)$. The two key outcomes we can learn from the simplified expression (\ref{Integral_analysis}) are the rectangular function and the shift in $\tau$ within the integral. The rectangular function has a width of DL and is centred such that the nonzero interval begins at 0. It physically corresponds to the time delays photons can acquire throughout the length of the crystal and ensures that the time delay distribution can only be nonzero in this specific interval. As for the specific shape of the time delay distribution, this depends on $g(\mathbf{q})$, which means that it is not predicted with this general analysis. The important thing, however, is that within the applicability of the approximations made, a change in the distance between collection lens and crystal, $z_{\mathrm{CL}}-z_{c}$, results in a shift of the time delay distributions, except for the fixed cut-off by the rectangular function. The shift is given by $nD\Delta z$, where $\Delta z$ is the displacement of the crystal or collection lens. It is important to bear in mind that this analysis applies to the case of fibre-coupled detection and to individual pairs of points on the detectors contributing to the free-space detection. 

We have evaluated the biphoton mode function and coincidences given by (\ref{SMFCoinc}) and (\ref{FSCoinc}) numerically without the approximations made later on, with the refractive indices modeled by temperature-dependent Sellmeier equations. A temperature dependence of the poling period caused by thermal expansion of the crystal was also taken into account based on coefficients from \cite{Pignatiello2007}, although this has a comparatively small impact. One of the detection schemes incorporates a bandpass filter, which we model with a Gaussian spectral transmission function applied to the biphoton wavefunction. The parameters required as inputs to the simulations for modelling the experiments are i. the fixed and known crystal length, poling period, and pump wavelength, ii. the variable but known pump beam waist and crystal temperature, and iii. the variable detection beam waist (for single-mode fibre detection) or detection area (for free-space detection). The detection beam waist was not measured experimentally but adjusted to optimise counts, and the free-space detection area is nominally $(\SI{50}{\micro m})^2$ but needed to be adjusted in the simulations to account for an imperfect imaging system.

To demonstrate the applicability of our analytic results from the toy model, which assumes a plane wave pump and is based on a Taylor approximation of the wavevector mismatch, to our experimental case of a focused pump beam, we present simulation results in figure \ref{fig:spatio-temporal-correlations} without the use of those approximations. Figure \ref{fig:spatio-temporal-correlations} (a) shows the spatio-temporal correlations for our set-up, with the linear dependence between $|\mathbf{q}|^2$ and $\Omega$ illustrated. It is interesting to note that this dependence differs from the case of type I down-conversion \cite{Gatti2009,Gatti2012,Gatti2014} due to the nonzero difference in the group velocities of signal and idler. The main outcomes from the simplified expression (\ref{Integral_analysis}) can be recognised in figure \ref{fig:spatio-temporal-correlations} (b), which shows time delay distributions for two different crystal positions using the single-mode fibre detection scheme. As discussed, the time delay distribution is shifted when the crystal is displaced, except for a fixed cut-off that remains and is modelled by the rectangular function. Of course unlike for a plane wave pump, the pump focal position comes into play for a focused pump. In our work we found that the focal position of the pump beam has little effect on the time delay shift, but has a significant impact on the proportion of photons detected, particularly depending on whether the pump and detection focal positions match up.

\section{Experimental Method} \label{Experimental Method}

\begin{figure}[ht]
	\centering
		\includegraphics[width=\textwidth]{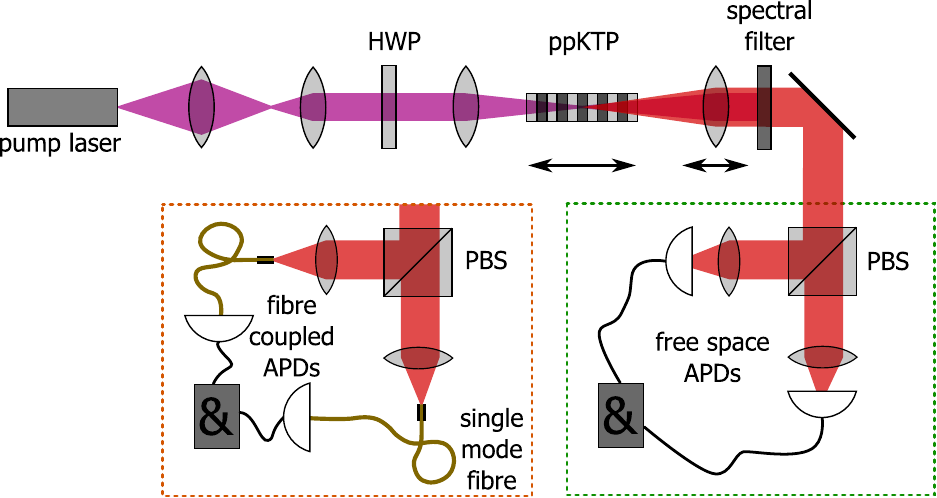}
	\caption{Schematic of the experimental set-up. The \SI{404.25}{nm} pump laser is expanded with a telescope to obtain a specific beam waist in the nonlinear crystal. A half waveplate (HWP) controls the polarisation relative to the crystallographic axes. A lens then focuses the collimated beam into the periodically poled potassium titanyl phosphate (ppKTP) crystal, from where the down-converted photons are collected by a collection lens. The pump beam is blocked by a longpass or interference bandpass filter. Signal and idler are separated by a polarising beam splitter (PBS) and subsequently either directly detected by two free-space avalanche photo diodes (APD) or first coupled to single-mode fibres and then detected. The time difference between signal and idler is measured with a time correlated single photon counting system.} 
	\label{fig:2014-06-12_TimeCorrelation_Paper_Exp_Setup}
\end{figure}

\begin{figure}[ht]
	\centering
		\includegraphics[width=0.6\textwidth]{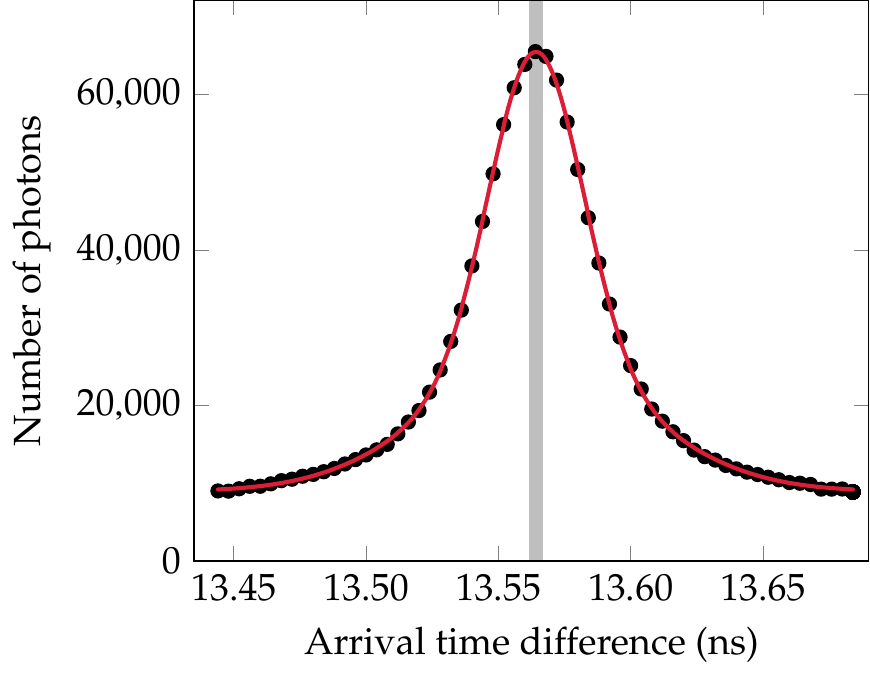}
	\caption{Example histogram showing the distribution of arrival time differences between signal and idler photons. Black circles indicate counted number of photons per time bin and the red line shows the fit to the data, from which the peak position is extracted. The width of the peak is dominated by the timing jitter of the APDs. Due to slight electrical differences between the two counting channels, there is always an offset in the time difference and only relative changes are important. The absolute time difference between signal and idler can be measured by exchanging them in the detection channels with a half waveplate. The integration time for the shown histogram was \SI{52}{s}. Each time bin is \SI{4}{ps} wide.  The temporal width over which the time delay of a photon pair from our crystal can in principle be varied is \SI{5.3}{ps}. This width is indicated as a grey stripe.} 
	\label{fig:2014-06-19_Fitted_Histogram}
\end{figure}

After describing theoretically the spatio-temporal correlations of the down-converted photons and their effect on the time delays in the detection of the photons, let us turn to describe our experimental apparatus and results. In figure \ref{fig:2014-06-12_TimeCorrelation_Paper_Exp_Setup} we show the experimental set-up used to create the SPDC photon pairs and how the time delay is measured. A continuous wave (CW), single frequency, diode laser with a wavelength of \SI{404.25}{nm} and a linewidth of approximately \SI{5}{MHz} is used to pump the \SI{15}{mm} long, periodically poled potassium titanyl phosphate (ppKTP) crystal in which the SPDC process takes place. The poling period of the crystal is \SI{9.89}{\micro m} to allow for type II down conversion at the pump wavelength.  The phase matching conditions fixing the central wavelength of the down-converted photons are controlled with the crystal temperature, which we set and keep to within a few \SI{10}{mK}. Control over the spatial modes of the down-converted photon pairs is possible by selecting the beam waist of the focused Gaussian pump beam inside the crystal \cite{Law2004}. We prepare the corresponding pump beam diameter by selecting a suitable telescope before the crystal. 

A longpass or a bandpass filter after the crystal blocks the pump beam. We measure the time delay between signal and idler photons with two different methods. Using free-space APDs (PDM SPD-050-CTE by Micro Photon Devices) we can collect a large portion of the down-converted field. The other option is to couple into single-mode fibres first and thus project into a Gaussian mode. 

As described earlier, the time delay in the photon detections is in the \si{ps} range. The timing jitter of standard Avalanche Photo Detectors is more than an order of magnitude larger, which renders a direct temporal characterization of the detection probability distribution very challenging. In our case, the distribution of arrival times is broadened by the timing jitter of both APDs to a full width at half maximum of about \SI{50}{ps}. On the other hand, the average time delay is amenable to be measured by carefully fitting the arrival time distribution. This measurement is only limited by the signal to noise in the measurements and the stability of the single photon counting system (PicoHarp 300 from PicoQuant). A typical histogram is shown in \mbox{figure \ref{fig:2014-06-19_Fitted_Histogram}}, together with an empirical fit that is used to extract the peak position and thus the average arrival time difference between signal and idler. The fitting model which was validated by the experimental data is based on a sum of two Gaussian profiles with different widths, positions and amplitudes, plus a constant background. This model contains just 7 free parameters from which we can extract the peak position. As the model fits very well the collected data, the error in the peak position is much smaller than the bin size. To further reduce the uncertainty, we acquire several histograms per crystal or collection lens position and average over the resulting peak positions. As a result we are able to measure average time differences much smaller than the time-binning of the histogram. To correct for a slow drift of the electronics, we acquire additional histograms at a reference position after each measurement. Both nonlinear crystal and the collection lens are mounted on motorized stages to scan their position and record the corresponding time difference.

\section{Experimental Results and Comparison with Theory}
\label{results}

\begin{figure}[ht]
	\centering
		\includegraphics[width=1\textwidth]{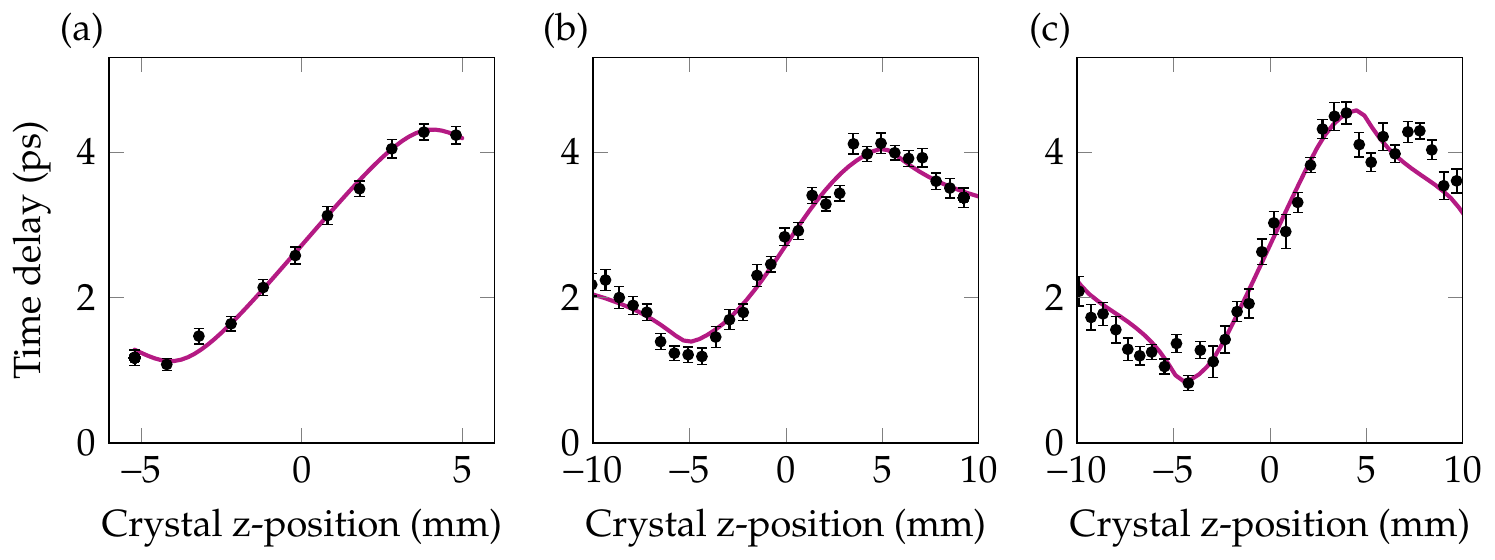}
	\caption{Time delay between signal and idler as a function of the crystal z-position. (a) With detection after incoupling into single-mode fibres, (b) with free-space detection after a \SI{2.5}{nm} wide bandpass filter, and (c) with free-space detection after a longpass filter. The purple line shows the theory prediction for the experimental parameters, which are: (a) $w=\SI{12.9}{\micro m}$, $T=\SI{59}{\degree C}$, $w_f=\SI{18}{\micro m}$, (b) $w=\SI{11.4}{\micro m}$, $T=\SI{58}{\degree C}$, $l=\SI{40}{\micro m}$, (c) $w=\SI{11.4}{\micro m}$, $T=\SI{60}{\degree C}$, $l=\SI{40}{\micro m}$. $l$ denotes the side length of the quadratic free-space detection area, which was adjusted in the simulations to account for an imperfect imaging system. The error bars span one standard deviation.}
	\label{fig:2014-06-16_TimeCorrelation_Crystal}
\end{figure}

The experimental results are presented in figure \ref{fig:2014-06-16_TimeCorrelation_Crystal} together with the results of the theoretical simulation described in section 2. The detection of the photons after coupling into single-mode fibres is shown in figure \ref{fig:2014-06-16_TimeCorrelation_Crystal} (a). We have also performed free-space detection with a spectral filter of width 2.5 nm (FWHM), figure \ref{fig:2014-06-16_TimeCorrelation_Crystal} (b), and without spectral filtering, figure \ref{fig:2014-06-16_TimeCorrelation_Crystal} (c). The simulation is based on the experimental parameters and the only adjustable parameter is the detection beam waist or detector size, depending on the type of measurement.

The results from different detection schemes are qualitatively similar. Each set of results shows a monotonic change in time delay within the range of crystal positions for which the detection focal point lies within the crystal, an interval of approximately \SI{8}{mm}. As mentioned above, in our convention the 0 crystal position is such that the pump and detection focal points coincide with the centre of the crystal. The range of time delays depicted in the figure corresponds to orthogonally polarised photons propagating a distance of up to the full crystal length in the birefringent crystal medium. Mathematically, these limits correspond to the cut-off by the rectangular function in equation (\ref{Integral_analysis}). A nonzero width in the distribution of time delays reduces the range of observed values since we are working with average time delays. The turning points visible in each case occur when the peak in the time delay distribution is shifted past the cut-off. Since we are measuring an average rather than the position of the maximum, shifting the peak past the cut-off results in the average moving back towards the central value. 

In addition to the dependencies shown, the specific shapes are further influenced by the detection beam waist for single-mode fibre coupling, the detector size for free-space measurement, and the crystal temperature. Spectral filtering is a common practice for a number of purposes such as decreasing the distinguishability of the two photons, and one of the other reasons is the  frequently desired consequence of reducing spatio-temporal correlations. However, as figure \ref{fig:2014-06-16_TimeCorrelation_Crystal} shows, the difference between the cases with and without spectral filtering is relatively minor, so in our case the reduction in the correlations was not sufficient to eliminate the mechanism. As mentioned previously, the distance between crystal and collection lens results in a particular time delay distribution. It is therefore also possible to achieve a change in time delay by moving the collection lens instead of the crystal. We have tried this in both experiment and theory and obtained results similar to those shown in figure \ref{fig:2014-06-16_TimeCorrelation_Crystal}. However, changing the crystal position is experimentally more relevant since this keeps the collimation of the beam after the collection lens intact.

\begin{figure}[ht]
	\centering
		\includegraphics[width=0.4\textwidth]{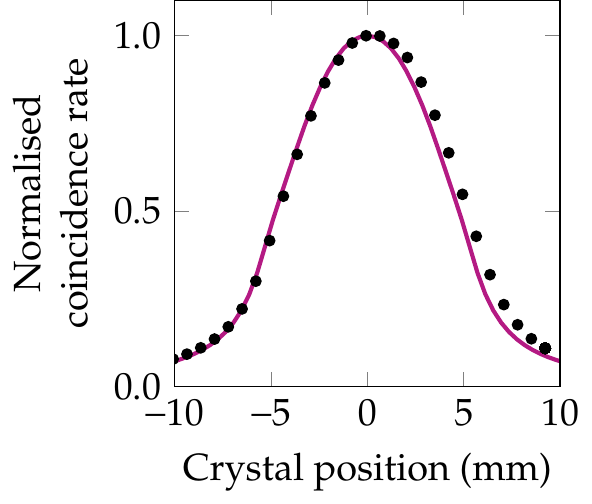}
	\caption{The coincidence count rate drops as the pump and detection foci move away from the centre of the crystal. Depicted are results for the free-space detection scheme with the \SI{2.5}{nm} wide bandpass filter. The purple line shows the theory prediction, where the same parameters were used as for figure \ref{fig:2014-06-16_TimeCorrelation_Crystal} (b).}
	\label{fig:2014-06-16_Coinc-Rate}
\end{figure}

It is important to note that even with such collimation and for free-space detection, not all
photons are detected and the coincidence rate is further reduced as the crystal
is moved from its central position. See figure \ref{fig:2014-06-16_Coinc-Rate} for
this behaviour, where we have measured and calculated the coincidence
rate in the experiment and the simulation shown in figure \ref{fig:2014-06-16_TimeCorrelation_Crystal} (b). The cause for this drop of the coincidence rate is due to a change in the spatial correlations of the photons when the crystal is moved, which causes the spatial distribution of the coincidences to be larger than the detector area. As can be seen in the figure, our theoretical model recovers the features of the coincidence rate quite faithfully.

\section{Discussion}

It is well known that SPDC is a coherent process and photons generated throughout the whole length of the crystal contribute to the biphoton wavefunction equally \cite{Search2003}. With this in mind, it is somewhat surprising to observe a change in time delay. The resolution lies in the fact that we do not detect all photons generated in the crystal. This is obvious for the case of single-mode fibre detection, where a significant portion of the down-converted field is rejected. Using a free-space detection scheme considerably increases the number of collected photons, but nonetheless, not all photons are collected due to the finite detector size. Indeed, in the simulations we only recover the experimentally measured change in time delay after restricting the collection of photons in space, compared to an infinite detector area. We believe such a loss of photons may also have caused the change of counts in \cite{Suzer2008}, from which the authors then infer a dependence of the SPDC efficiency on the pump beam intensity. Changing the distance between crystal and collection lens is modeled by the phase term $\mathrm{exp}\left(id \left(\frac{|\mathbf{q}_{s}|^2 }{2 k_{air}(-\Omega) }+\frac{|\mathbf{q}_{i}|^2 }{2 k_{air}(\Omega) }\right)\right)$ where $d=\left(f_1-\left(z_{\mathrm{CL}}-z_c-\frac{L}{2}\right)\right)$. This term influences the time delay due to spatio-temporal correlations.

The time delay between photons plays an important role in quantum interference experiments such as the one by Hong, Ou, and Mandel \cite{Hong1987}. In the case where the photons travel along separate paths before incidence on the beam splitter, it is easy to control the path length difference over a wide range. In contrast, a Hong-Ou-Mandel experiment in which signal and idler photons travel along the same path before arriving at a polarising beam splitter \cite{Resch2001,Kuklewicz2004}, requires birefringent materials to control the time delay. When using a fixed delay compensation, our results show that it is important to have the crystal and collection lens positioned correctly, even if spectral filtering comparable to our 2.5 nm bandpass filter is performed. This is particularly relevant for the increasingly popular use of long crystals for bright sources of photon pairs, made possible through the use of periodic poling. Conversely, our observed dependence could also be harnessed by controlling the time delay through a deliberate positioning of the crystal. It is important to keep in mind, however, that the change in time delay achieved in this way is not equivalent to a shift in the complete time delay distribution, as is the case with a relative difference in free-space propagation lengths, and that it is influenced by the restricted detection. The restriction of the detection depends on the ratio of the focal lengths of the collection lens and the lenses before the detectors, and on the detector area.

It is interesting to compare our findings with reference \cite{Jedrkiewicz2012}. Similarly to our work, a quadratic phase associated with the displacement of an experimental component was shown to be responsible for a change of the time delay distribution. However, since the type I phase matching in that case entails different spatio-temporal correlations, a dispersion like effect instead of a time delay shift was observed.

Throughout this work, we have presented results of average time delays. Experimentally, this was necessary because of the timing jitter of the detectors, which is much larger than the range of observed changes. In future experiments, it would be interesting to extract more information about the time delay distribution.

\section{Conclusion}

We have shown for a number of experimentally relevant detection schemes, that the average time delay between signal and idler photons in type II collinear down-conversion depends on the distance between the nonlinear crystal and the collection lens. The possible change of time delay covers a large portion of the total delay photon pairs can acquire by propagating through the whole length of the crystal. We have measured the time delay for single-mode fibre detection, as well as free-space detection with and without spectral filtering. The experimental results are well described by our model. The reason for the change in time delay are spatio-temporal correlations and the selective nature of the detection, even in the case of free-space measurements. On the one hand, the observed effect is something to beware of when indistinguishable photons are required. On the other hand, when used deliberately, this constitutes a novel way to tune the delay.

\section*{References}
\bibliography{References}
\end{document}